\documentclass{aa}  

\usepackage{graphicx}
\usepackage{epsfig}
\usepackage{epstopdf}
\usepackage{txfonts}
\usepackage{natbib}
\newcommand{\raiseChi}

\begin{document}

   \title{Unveiling the gas and dust disk structure in HD 163296 using ALMA observations}


   \author{I. de Gregorio-Monsalvo\inst{\ref{inst1},\ref{inst6}}\and F. M\'enard\inst{\ref{inst2},\ref{inst3}}\and W. Dent\inst{\ref{inst1},\ref{inst10}}\and C. Pinte\inst{\ref{inst3}}\and C. L\'{o}pez\inst{\ref{inst1}}\and P. Klaassen\inst{\ref{inst4}}\and A. Hales\inst{\ref{inst1}}\and P. Cort\'es\inst{\ref{inst1}}\and M. G. Rawlings\inst{\ref{inst5}}\and K. Tachihara\inst{\ref{inst1},\ref{inst9}}\and L. Testi\inst{\ref{inst6},\ref{inst7}}\and S. Takahashi\inst{\ref{inst8}}\and E. Chapillon\inst{\ref{inst8}}\and G. Mathews\inst{\ref{inst4}}\and A. Juhasz\inst{\ref{inst4}}\and E. Akiyama\inst{\ref{inst9}}\and A. E. Higuchi\inst{\ref{inst1},\ref{inst9}}\and M. Saito\inst{\ref{inst1},\ref{inst9}}\and L.- \AA. Nyman\inst{\ref{inst1},\ref{inst10}}\and N. Phillips\inst{\ref{inst1},\ref{inst10}}\and  J. Rod\'on\inst{\ref{inst10}} \and S. Corder\inst{\ref{inst1}} \and T. Van Kempen\inst{\ref{inst1},\ref{inst4}}
          }

   \institute{Joint ALMA Observatory (JAO), Alonso de C\'ordova 3107, Vitacura, Santiago de Chile. \email{idegrego@alma.cl}\label{inst1}
\and
   UMI-FCA, CNRS/INSU France (UMI 3386), and Departamento de Astronom\'{\i}a, Universidad de Chile, Santiago, Chile.\label{inst2}
\and
  UJF-Grenoble 1/ CNRS-INSU, Institut de Plan\'etologie et d'Astrophysique de Grenoble (IPAG) UMR 5274, Grenoble, France\label{inst3}
 \and
   Leiden Observatory, Leiden University, PO Box 9513, 2300 RA Leiden, the Netherlands\label{inst4}
 \and
   National Radio Astronomical Observatory (NRAO), 520 Edgemont Road, Charlottesville, VA 22903, USA\label{inst5}
 \and
   European Southern Observatory, Karl Schwarzschild Str 2, D-85748 Garching bei M\"unchen, Germany\label{inst6}
 \and
   INAF-Osservatorio Astrofisico di Arcetri, Largo E. Fermi 5, I-50125 Firenze, Italy\label{inst7}
 \and
  Academia Sinica Institute of Astronomy and Astrophysics, P.O. Box 23-141, Taipei 10617, Taiwan\label{inst8}
 \and
  National Astronomical Observatory of Japan (NAOJ), 2-21-1 Osawa, Mitaka, Tokyo 181-8588, Japan\label{inst9}  
\and
  European Southern Observatory, Alonso de C\'ordova 3107, Vitacura, Santiago, Chile\label{inst10}  
        }


   \date{Accepted for publication in A\&A}

 
  \abstract
   {}
   {The aim of this work is to study the structure of the protoplanetary disk surrounding the Herbig Ae star HD 163296.}
   {We have used  high-resolution and high-sensitivity ALMA observations of the ${\rm CO}(3$--$2)$ emission line and the continuum at $850\,\mu{\rm m}$, as well as the 3- dimensional radiative transfer code MCFOST to model the data presented in this work.}
   {The ${\rm CO}(3$--$2)$ emission unveils for the first time at sub-millimeter frequencies the vertical structure details of a gaseous disk in Keplerian rotation, showing the back- and the front-side of a flared disk. Continuum emission at $850\,\mu{\rm m}$ reveals a compact dust disk with a $240\,{\rm AU}$ outer radius and a surface brightness profile that shows a very steep decline at radius larger than $125\,{\rm AU}$. The gaseous disk is more than two times larger than the dust disk, with a similar critical radius but with a shallower radial profile. Radiative transfer models of the continuum data confirms the need for a sharp outer edge to the dust disk. The models for the ${\rm CO}(3$--$2)$ channel map require the disk to be slightly more geometrically thick than previous models suggested, and that the temperature at which CO gas becomes depleted (frozen-out) from the outer regions of the disk midplane is $T<20\,{\rm K}$, in agreement with previous studies.}
   {}
      
   \authorrunning{de Gregorio-Monsalvo et al.}
   \titlerunning{ALMA observations of HD 163296 disk structure} 
   \maketitle
%

\section{Introduction}

Dust and gas-rich disks around recently-formed stars are important as they harbor planets either recently or possibly still in the process of forming.  The structure of these young protoplanetary disks has been the subject of intense study over a wide range of wavelengths (see \citealt{Wil11} for a recent review).   Their sizes range up to a few hundred AU with Keplerian rotation velocities and temperatures up to a few tens of K in the disk midplane several AUs from the central star. This means that millimeter/sub-millimeter telescopes are well suited to study their molecular and dust components. Single-dish telescopes generally cannot resolve the disks, but have revealed the characteristic double-peaked line profiles from CO and several other species. Interferometers at millimeter wavelengths have provided images a few resolution elements across, but hitherto do not provide highly detailed images.

At a distance of $122\,{\rm pc}$, \object{HD 163296} is an isolated A2Ve star of estimated ${\rm age}\approx5\,{\rm Myr}$ \citep{Mon09}. It is one of the best-studied protoplanetary disks, and was one of the first to be resolved with millimeter interferometry \citep{Man97}.  Sub-millimeter interferometry observations with 2$\arcsec$ resolution indicates an outer radius of $550\,{\rm AU}$ in CO, with an inclination of $45^\circ$ \citep{Ise07}. The brightness of this source in millimeter molecular lines has made HD 163296 an excellent laboratory for comparing with disk models, provoking studies of radial and vertical temperature and molecular abundances \citep{Qi11,Til12,Aki11}. Determining unique solutions to the underlying disk structure clearly benefits from having high-angular resolution, good image fidelity and high-sensitivity. Detailed studies of dust and gas structures in protoplanetary disks are now possible with the spatial resolution and sensitivity provided by ALMA. 

In this work we present a detailed study of the disk structure surrounding HD 163296 using the best images to date provided by ALMA band 7 data in continuum and spectral line.


\section{Observations description}

Observations were performed on 2012 June 9, 11, 22, and July 6 at Band 7, as part of the ALMA science verification program 2011.0.000010.SV. The array was in a configuration with projected baselines length between $\sim$16 to $\sim$400 m, sensitive to maximum angular scales of $\sim\!\!7''$ and providing a synthesized beam of $0.52\arcsec \times 0.38\arcsec$ at ${\rm P.A.}\!\sim\!\!82^\circ$. The field of view was $\sim\!\!18''$. A total of five data sets were collected, using between 16 and 19 antennas of $12\,{\rm m}$ diameter and accounting for 3.9 hours of total integration time including overheads and calibration (2.3 hours on HD 163296).   Weather conditions were good and stable,  with an average precipitable water vapor of 0.8 mm. The system temperature varied from 100 to $300\,{\rm K}$. 

The correlator was set up to four spectral windows in dual polarization mode, centered at 345.796 GHz (${\rm CO}(3$--$2)$), 346.998 GHz (${\rm H^{13}CO}(4$--$3)$), 356.734 GHz (${\rm HCO^{+}} (4$--$3)$), and 360.170 GHz (${\rm DCO^{+}} (5$--$4)$). The effective bandwidths used were $468.75$, $937.50$, $468.75$, and $117.19\,{\rm MHz}$, respectively, at velocity resolutions of  $\sim$ 0.21,  0.42,  0.21, and 0.05 km s$^{-1}$ after Hanning smoothing.  In this letter we present results from the observations in the ${\rm CO}(3$--$2)$ emission line and the continuum at $\sim$$850\,\mu{\rm m}$.

The ALMA calibration includes simultaneous observations of the $183\,{\rm GHz}$ water line with water vapor radiometers, which measure the water column in the antenna beam, later used to reduce the atmospheric phase noise.   Amplitude calibration was done using Neptune, and quasars J$1924$$-292$ and J$1733$$-130$ were used to calibrate the bandpass and the complex gain fluctuations respectively.
Data reduction was performed using the version 3.4 of the Common Astronomy Software Applications package (CASA). We applied self-calibration using the continuum and we used the task CLEAN for imaging the self-calibrated visibilities. The continuum image was produced by combining all of the line-free channels. 
 Briggs weighting was used in both continuum and line images. The achieved rms was 0.5 mJy beam$^{-1}$ for the continuum and 14 mJy beam$^{-1}$ for each CO channel map. 

\section{Results and modelling \label{Results}}
\subsection{ Continuum emission at $850\,\mu{\rm m}$ \label{Continuum}}

\begin{figure}
\begin{center}
\includegraphics[width=10cm]{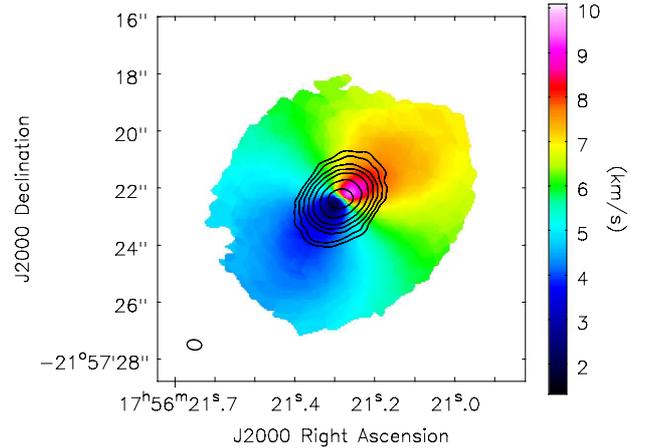}
\caption{Colours represent the ${\rm CO}(3$--$2)$ integrated emission over all velocity channels (first-order moment, 5$\sigma$ cut). Contours show the continuum emission at $850\,\mu{\rm m}$ with levels representing 5, 10, 20, 40, 80, 160, and 320 times the rms of the continuum map ($0.5\,{\rm mJy}~{\rm beam^{-1}}$). }\label{fig1}
\end{center}
\end{figure}

Continuum emission at $850\,\mu{\rm m}$ is detected centered at the position R.A.(J2000) = 17$^{h}$56$^{m}$21$\fs$285, Dec(J2000) =$-$21$\degr$57$\arcmin$22$\arcsec$368. The dusty disk is well resolved with no suggestion of gaps or holes at radius $>\!25\,{\rm AU}$.  The major axis has a projected diameter of $3.9''$ (measured at the 3$\sigma$ level), which corresponds to an outer dust disk radius of $R_{\rm out}\!\simeq240\,{\rm AU}$ at the adopted distance of $122\,{\rm pc}$ \citep{vda98}. This value is $\sim$20$\%$ larger than the one reported by \cite{Ise07}  and can be explained by the higher sensitivity and resolution of the ALMA data, unveiling the fainter outer edge of the dusty disk (see Fig~\ref{fig1}).  The inclination angle is $i=45^\circ$, derived from the shape of the isophote contours, and it is in agreement with \cite{Ise07}. The major axis position angle is 137$\degr$. The flux density integrated over all the disk structure above 3$\sigma$ is  $S_{850\mu m}$= 1.82 $\pm$0.09 Jy, similar to the values reported by \cite{Ise07} and \cite{Qi11} using Submillimeter Array (SMA) observations at similar wavelengths.  

The continuum surface brightness profile was calculated by azimuthally averaging the emission from concentric elliptical annuli from the central star. It can be fitted by a two-component power law, with a characteristic radius ($R_{C}$, outside of which the brightness profile drops toward zero) equal to $125\pm5\,{\rm AU}$ from the central star, changing from a power-law slope of $1.03\pm0.04$ to a very steep decline of $4.7\pm0.2$ (see Fig~\ref{rad}). A minimized {\raise0.3ex\hbox{$\chi$}}$^{2}$ fitting was used to determine the best value of the power-law slopes. 

\subsection{${\rm CO}(3$--$2)$ spectral line emission \label{Line}}

The CO data show a disk in  Keplerian rotation (see Fig~\ref{fig1}). Considering contours above 3$\sigma$, the outer radius extends to $R_{\rm out}\!\simeq575\,{\rm AU}$, in agreement with \cite{Ise07}. The integrated intensity averaged over the whole emission area is $100.30\pm0.13$ Jy km s$^{-1}$, in agreement with the measurement reported by \cite{Qi11}. The inclination and the P.A. of the gaseous disk (measured at the 3$\sigma$ contour level) are 38$\degr$ and 138$\degr$ respectively.  We note the inclination value is different from the one using the continuum due to optical thickness and flaring combined with high-spatial resolution effects (see next section). For our analysis we adopt the values calculated from the continuum, less affected by the flared structure of the disk. 

The same fitting as done for the continuum in \S\ref{Continuum} was applied to the line data. In this case the surface brightness profile of the ${\rm CO}(3$--$2)$ can also be approximated by a power law with a break in the surface brightness at $125\pm5\,{\rm AU}$ (similar to that of the continuum). At radii $125$--$240\,{\rm AU}$, the CO surface brightness drops with a power law of $1.0\pm0.1$ and beyond $240\,{\rm AU}$, the CO drops rapidly, with a power law of $2.5\pm0.1$ (considerably less steeply than the continuum; see Fig~\ref{rad}).

\subsection{Modelling \label{Modelling}}

To extend beyond the simple power law fitting, the 3-dimensional radiative transfer code MCFOST was used to model the data presented in this work (see \citealt{Pin06,Pin09} for a more detailed description). The initial disk model we use is based on \cite{Til12}, in particular their favored model number three (see Table 6 in that paper). The same photospheric parameters were used, namely $T_{eff}$ = 9250~K, $M_{*}$ = 2.47 $M_{\odot}$, $L_{*}$ = 37.7~L$_{\odot}$, and a slight UV excess, L$_{UV}/ L_* = 0.097$. The model contains an exponentially tapered-edge and provided an adequate fit to the broadband spectral energy distribution (SED), the Herschel PACS lines observed by the Key program GASPS (Gas in Protoplanetary Systems; \citealt{Den13}), as well as sub-millimeter interferometry (continuum and CO data, from \citealt{Ise07}). 

The temperature structure in the disk is calculated by considering the dust opacity only, assuming astronomical silicates \citep{Dra84} with a power-law size distribution having a slope of $-3.5$ and a maximum radius of 1~mm. Each cell in the computation domain has its own density, grain properties, and opacity constructed following the global disk parameters. The calculated dust temperature profile ($T_{\rm dust}$) decreases outward and there is a vertical temperature gradient for the dust that also depends on radius, with the midplane being cooler than the disk surface. 

To calculate the ${\rm CO}(3$--$2)$ channel maps and surface brightness distribution MCFOST assumes a constant gas-to-dust mass ratio of 100 throughout the disk (both radially and vertically). We adopted a standard CO abundance with respect to H$_2$ (10$^{-4}$), set constant through the disk where $T_{\rm dust}>20\,{\rm K}$ and equal to zero where $T_{\rm dust}<20\,{\rm K}$  to mimic the effect of CO freeze-out (see \S\ref{vertical}).  The level populations are calculated assuming LTE and $T_{\rm gas}(r,z)=T_{\rm dust}(r,z)$ for each grid cell.
The radial and vertical temperature profiles and the radiation field estimated by the Monte Carlo simulation are used to calculate level populations for the CO molecule and to produce the SED, continuum images, and line emission surface brightness profiles and kinematics with a ray-tracing method. The kinematics are calculated  assuming the disk is in pure Keplerian rotation. 

\section{Discussion \label{Discussion}}

\subsection{Dust and gas surface brightness distributions\label{surface}}

In order to fit the brightness radial profiles observed in the continuum at $850\,\mu{\rm m}$ and in the ${\rm CO}(3$--$2)$ emission line, we consider a tapered-edge model for the surface density distribution (see \citealt{And09}): 
\begin{equation}
 \Sigma  = \Sigma_{c}\left(\frac{R}{R_{c}}\right)^{-\gamma} \exp \left[-\left(\frac{R}{R_{c}}\right)\right]^{2-\gamma}
\end{equation}
\noindent where $R_{\rm c}$ is the characteristic radius and $\gamma$ is the index of the surface density gradient.  In our data, CO is detected over a radius more than twice the dust continuum radius. This feature, for which the tapered-edge disk model has provided a solution for previous studies at lower spatial resolution and sensitivity, was produced because the CO line opacity is very much larger than the dust continuum opacity at $850\,\mu{\rm m}$, such that the CO gas remains optically thick and detectable over a much larger radius. 

\subsubsection{The outer disk, outside of $R_{\rm c}$ \label{out}}

 The initial disk model was based on the best model found by \cite{Til12}, which adequately fits to the broadband SED, the [OI] 63 $\mu$m Herschel PACS line, as well as previous sub-millimeter interferometric observations in  ${\rm CO}(3$--$2)$,  ${\rm CO}(2$--$1)$, and  $^{13}{\rm CO}(1$--$0)$.  For performing the fitting to the radial profiles of the ${\rm CO}(3$--$2)$ and the continuum, the $R_{\rm c}$ and the power law slope derived from the simple initial fitting described in \S\ref{Results} were used as a starting point. The previous value of $\gamma$ = 1, reported in the literature by \cite{Hug08} for simultaneously fitting the continuum and the ${\rm CO}(3$--$2)$ emission as observed with the SMA, was also considered as a reference for the fitting. A range of $\gamma$ values from 0 to 2 were then explored to reproduce the steep decline in continuum and the shallower ${\rm CO}(3$--$2)$ surface brightness profile (see Table~\ref{grid} for a description on the parameter space used). 

Good model values for $R_{\rm c}$ and $\gamma$ were found separately for the continuum and for the line surface brightness data.  No minimisation scheme was used, but we note that the continuum model and the line model (outside of $R_{\rm c}$ ) match all the data point within the error bars (see discussion below for the inner $100\,{\rm AU}$ in the CO line). A posteriori, we verify that both models use the same value for $R_{\rm c}$, however the line and continuum require very different values for $\gamma$. After the best $\gamma$ value was found, the total flux density and the size of the emission provided by the model were compared with the values obtained observationally for both continuum and line. A dust mass of $7\!\times\!10^{-4}M_\odot$ \citep{Til12} was assumed and a comparison of the synthetic SED with the observations was performed also for consistency.

\begin{figure}[t!]
\includegraphics[width=.7\textwidth, angle=0, scale=0.65]{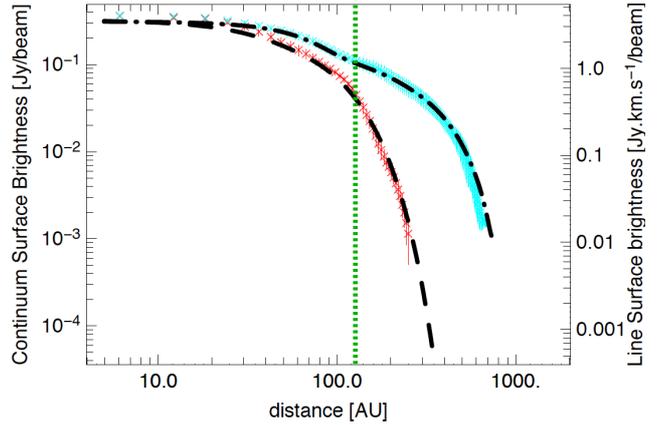}
\caption{Surface brightness profiles for the continuum (red crosses) and for the ${\rm CO}(3$--$2)$ (light blue crosses) with 1$\sigma$ error bars. Dashed and dashed-dotted lines represent two different model fits required for the continuum (with $\gamma=0.1$) and for the spectral line ($\gamma=0.9$) profiles respectively. Green dotted line marks $R_{\rm c}$=125 AU}\label{rad}
\end{figure}

As a result, the best match for the very steep decline of the continuum surface brightness outside $R_{\rm c}$ requires very small values of $\gamma$  ($<0.1$; see Figure~\ref{rad}), reproducing the total integrated continuum flux well (5$\%$ larger than the observed integrated flux), the $R_{\rm c}$, and the radial profile of the surface brightness. But such a small gamma cannot fit the integrated ${\rm CO}(3$--$2)$ radial profile and the model line emission would be only detectable out to $r\sim400\,{\rm AU}$, while it is observed out to $575\,{\rm AU}$.
Using  $\gamma=0.8$--$0.9$ provides a better match to the ${\rm CO}(3$--$2)$ surface brightness profile data, as well as for integrated CO flux ($1\%$ higher than the observed one),  and $R_{\rm c}$ (see Fig~\ref{rad}). However, a value of $\gamma$ as high as this would result in a continuum disk whose surface brightness decline is too shallow.  Values of $\gamma$ near $1.0$ have been reported in previous studies at lower resolution to fit well both the spectral line emission and the continuum  (e.g., \citealt{Hug08}).  Those values are ruled out for the continuum by the higher angular resolution data reported here.

\begin{figure*}[t!]
\centering
\includegraphics[width=18cm,clip]{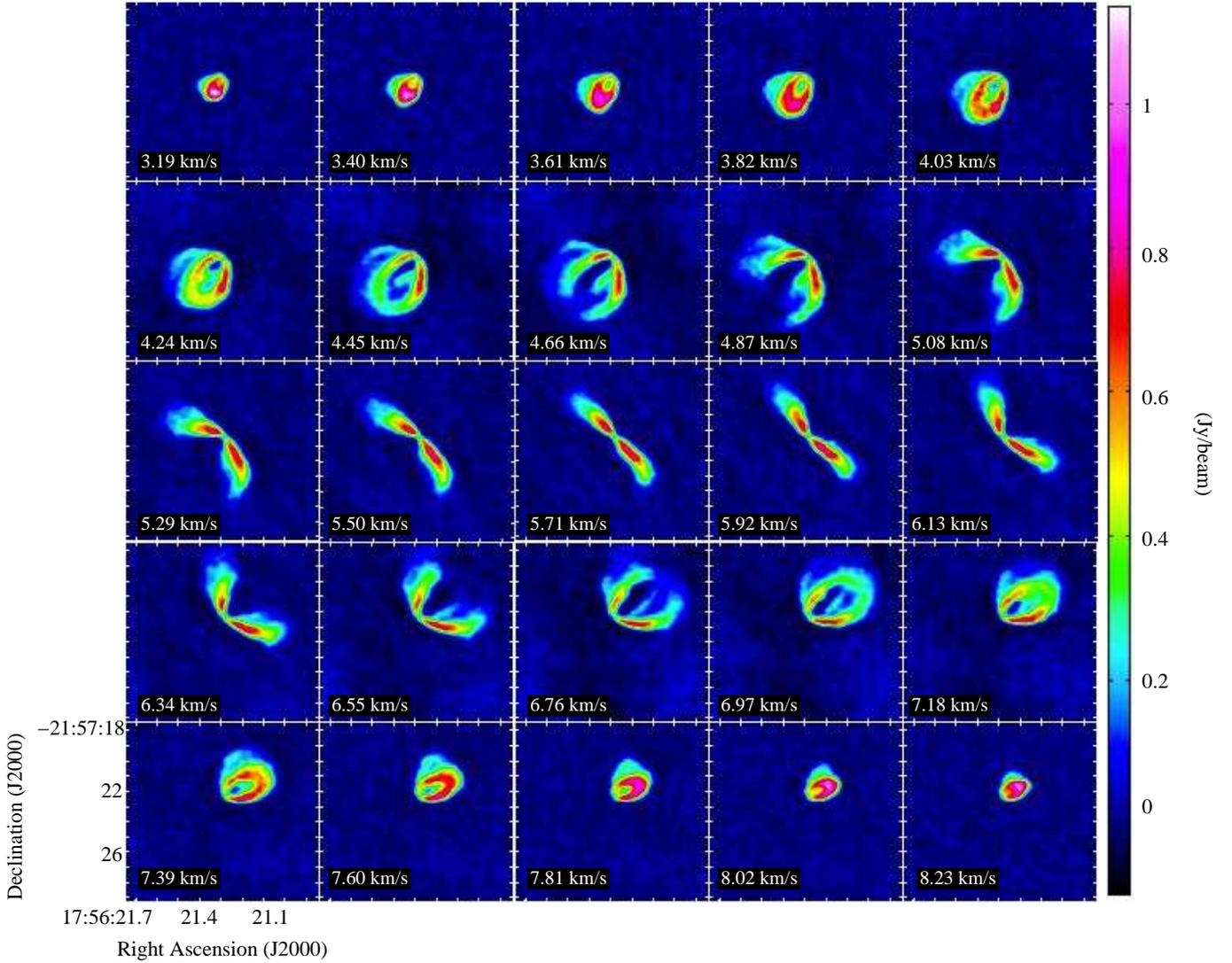}
\caption{Channel maps of the ${\rm CO}(3$--$2)$ emission in HD 163296 from $3.19$ to $8.23\,{\rm km}~{\rm s ^{-1}}$. The spectral resolution is $0.21\,{\rm km}~{\rm s ^{-1}}$ and the rms per channel is $14\,{\rm mJy}~{\rm beam^{-1}}$.}
\label{appfig}
\end{figure*}

\subsubsection{The central region of the disk, inside of $R_{\rm c}$ \label{int}}

For the continuum data, the same model with small $\gamma$ matches the observations nicely and no modifications are required. However, there is an extra component for the CO line emission at the center of the disk that is visible in Fig.~\ref{rad}. It shows also as a break in the ${\rm CO}(3$--$2)$ surface brightness profile just inside of $R_{\rm c}$. 
In order to investigate the origin of this extra emission, we considered further the results of \cite{Til12} from which our model is inspired. They showed (see their Figure 7, right panel) that there is a layer of hot gas at the surface of the disk, significantly hotter than the dust, for radii smaller than $50$--$80\,{\rm AU}$.  MCFOST does not calculate the gas temperature, but we can verify easily that the model CO surface brightness profile is a good match to the data when using $T_{\rm gas}=1.5T_{\rm dust}$ for a radius $<\!80\,{\rm AU}$, and $T_{\rm gas}=T_{\rm dust}$ elsewhere. This is the model shown in Fig.~\ref{rad}, where the resulting change in the brightness profile, after convolution by the synthesized beam, is shown to match the observed profile. In this slightly modified model, the same value of gamma = 0.9 is used for the CO gas disk, and the extra emission is coming only from the increased in gas temperature at small radii, with no impact on the outer disk surface brightness profile.

\begin{figure*}
\includegraphics[angle=0,scale=.73]{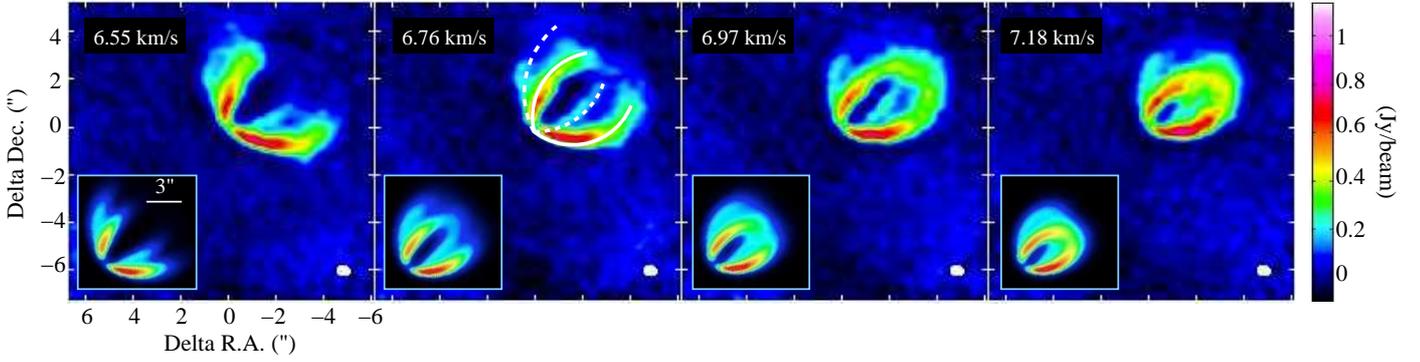}
\caption{${\rm CO}(3$--$2)$ representative channel emission maps in HD 163296, primary beam corrected.  The spectral resolution is 0.21 km s$^{-1}$ and the rms per channel is 14 mJy beam$^{-1}$.  The synthesized beam size is represented in the lower-right in each panel. The white solid and dotted lines represent the front and the back gas disk surface respectively. The insert at bottom-left of each panel shows a zoomed-down model at the corresponding velocity (i.e. the size is the same, the scale is different).}\label{channels}
\end{figure*}

\begin{table*}
\small
\caption{\label{grid} Parameter space explored by the presented analysis}
\centering
\begin{tabular}{llcll}
\hline\hline
Parameter & Description & Min-Max  &Step  &Best value \\
\hline
\hline

$R_{c}$  [AU]                                    & Critical Radius                         & $110-140$       & 5           & 125  \\
$\gamma$                               &Index of the surface density gradient   & $0-2$       &0.1      &0.1, 0.9\tablefootmark{a}\\
$T_{CO fr}$  [K]                       &CO freeze-out temperature               & $10-40$       &5        &20\\
$dV_{turb}$ [${\rm km}~{\rm s ^{-1}}$]   &Turbulent velocity width                & $0-0.3$       &0.05     &0.1 \\
$H_{0}$ at 1 AU [AU]                    &Scale height                            & $0.03-0.1$    &0.01     &0.07  \\
$\beta$                                 &Flaring angle index                     & $1.0-1.21$    &0.06     &1.12  \\

\hline
\hline                                                        
\end{tabular}                                                 
\tablefoot{
\tablefoottext{a}{$\gamma = 0.1$ is the best value for the continuum data, and $\gamma = 0.9$ for the CO spectral line data.}
}
\end{table*}

\subsection{Origin of the different surface brightness profiles \label{Origin}}
 Coronagraphic imaging with the Hubble Space Telescope toward HD 163296 revealed a scattered light disk with an outer radius that extends to at least $450\,{\rm AU}$ \citep{Gra00}. This suggests sub-micron dust particles responsible for scattered light remain coupled to the gas,  while larger particles ($\sim\!100\,\mu{\rm m}$ or larger) responsible for the $850\,\mu{\rm m}$ emission are concentrated in a smaller radius closer to the central star. One plausible explanation for the clear difference in the radial distributions of millimeter-sized dust grains and CO gas is a combination of grain growth and inward migration.  It is known that grain growth occurs in this disk, as reported by \cite{Ise07} based on the slope of the dust opacity law in the interval $0.87$-$7\,{\rm mm}$. Related to the inward migration,  models predict  the gas drag to be more efficient in intermediate size dust particles, with small grains remaining coupled to the gas (as seen in scattered light) and boulders following marginally perturbed Keplerian orbits (e.g. \citealt{Bar05}). Models predict a sharp cut-off of the sub-millimeter continuum emission when both grain growth and inward migration are considered (e.g. \citealt{Lai08}), in agreement with our observations. 
A similar case of dusty compact disk with a sharp outer edge smaller than the CO gaseous disk has been reported by \cite{And12}  for \object{TW Hya}.

Radial dust migration, expected to occur in disks massively, is an important ingredient for the evolution of disks and the formation of planets. Nevertheless it remains a phenomenon that is poorly constrained by observations. More efforts are needed to identify new sources like TW Hya and HD 163296 to better understand and characterise the physical processes leading to these more compact continuum disks at sub-millimeter wavelengths.

\subsection{The vertical structure of HD 163296 \label{vertical}}
The ${\rm CO}(3$--$2)$ channel maps unveil clearly for the first time at sub-millimeter frequencies the vertical structure of a flared gaseous disk (see Fig~\ref{appfig}). In Fig~\ref{channels} we have plotted four representative channel maps that show bright CO emission from the front disk surface, as well as similar but fainter and apparently rotated emission from the rear disk surface.   This ``two-layer'' effect was predicted theoretically by \cite{Sem08} and it arises because the disk is tilted, the ${\rm CO}(3$--$2)$ is optically thick and its emission is close to the disk surface, where the flaring effect is more prominent. 

In order to reproduce the spatial and kinematic pattern observed in the CO channel maps the CO gas model derived in the previous section was used, with $\gamma=0.9$ and without the extra heated CO layer in the center. Because there is an intrinsic vertical temperature gradient calculated in the disk (with the midplane being colder) ”two-layer” channel maps are produced naturally, with lower intensity CO emission coming from the midplane compared to the surfaces. However, without further changes to the model, these ”two-layer” channel maps do not produce enough contrast, enough difference in surface brightness between the midplane and the surface layers, even when convolved by the proper beam. Also, the two surface layers appeared slightly too close to one another. Both effects are suggesting the need for a geometrically thicker disk and/or a larger vertical temperature gradient.


In order to improve the match between modeled and observed channel maps, we fine-tuned the CO model slightly by exploring a set of values of CO freeze-out temperature, non-thermal turbulence velocity, scale height and flaring power. A summary of the parameter space explored is shown in Table~\ref{grid}.

First, we analyzed the effect of the variation of the freeze-out temperature in the patterns in the channel maps. The contrast between the warmer upper layers and the colder midplane can be increased by directly removing CO gas from the midplane. This can be done for example when the temperature is low enough that CO ice is produced, removing CO from the gas phase. To mimic the effect of freeze-out in the midplane, we set the CO gas abundance to zero wherever $T_{\rm gas}=T_{\rm dust}$ is below a critical value. Based on the initial value of 19 K adopted by \cite{Qi11}, a range of higher and lower values were explored.  In particular a range of freeze-out temperatures from 10 to $40\,{\rm K}$ were examined. The best match is obtained for $20\,{\rm K}$,  similar to the value obtained by \cite{Qi11}.  For $20\,{\rm K}$, the layer where CO is removed  in our model has a  vertical extend of $15\,{\rm AU}$ (at $200\,{\rm AU}$ radius), and it affects mostly the outer disk midplane. This provides direct evidence for CO freeze-out close to the disk midplane in HD 163296 (see also Mathews et al. 2013 submitted).  Freezing-out CO gas improves the match between the model and the observations.

Then the effect of turbulence was also estimated by varying the non-thermal velocity component of the line width. Values similar to those estimated in other disks (e.g. \citealt{Hug11}) were considered. In particular, values between $0$ to $0.3\,{\rm km}~{\rm s ^{-1}}$ were explored and compared with the spatial extent of the emission in a given channel. The best match is obtained for $0.1\,{\rm km}~{\rm s ^{-1}}$.  This is slightly less than the value of  $0.3\,{\rm km}~{\rm s ^{-1}}$  suggested for the upper layers of the disk in HD 163296 by  \cite{Hug11} for the CO(3-2) line as well, although a significant uncertainty is attached to this number. In our calculations, values higher than  $0.2\,{\rm km}~{\rm s ^{-1}}$ smear out the emission patterns too much in each channel.

The apparent separation between the two layers of CO emission in the channel maps is a function of the system's inclination and geometry (thickness) of the disk. The inclination is well known, as derived from the observations. The geometry of the disk in our model is defined in a large part by the reference scale height ($H_0$) and the flaring exponent. Starting from  the initial scale height and flaring index reported by \cite{Til12}, a range of higher and lower values of the scale height (between $0.03$ and $0.1$ at $1\,{\rm AU}$) and flaring index (from $1.00$ to $1.21$) were considered (see Table~\ref{grid}). A good match is obtained for a scale height of $0.07\,{\rm AU}$ (at $1\,{\rm AU}$ radius), and a flaring power of 1.12. These values correspond to a disk that is slightly geometrically thicker than the disk of \cite{Til12}.

\begin{figure}[t!]
\includegraphics[width=.7\textwidth, angle=0, scale=0.65]{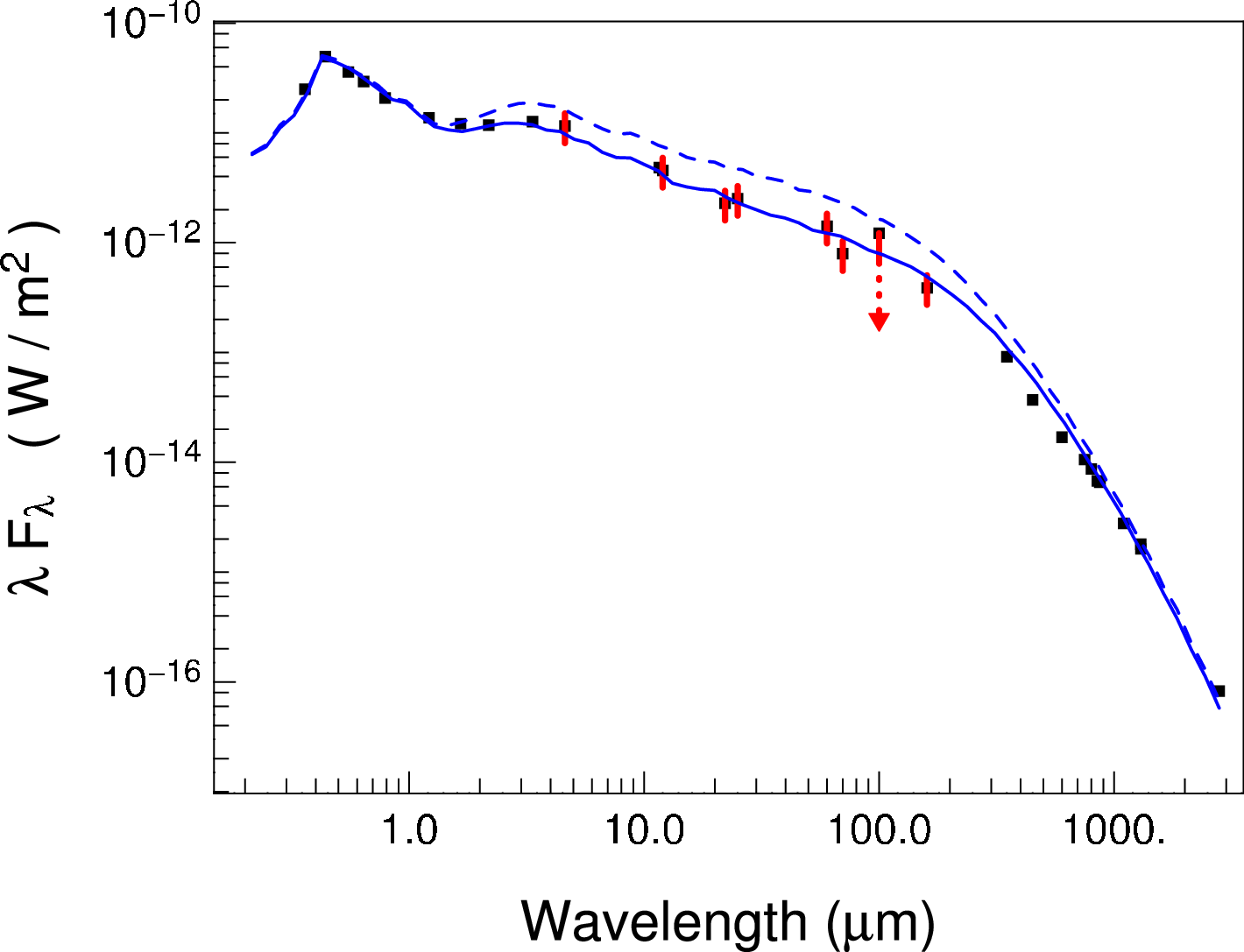}
\caption{Spectral energy distribution of HD163296. Photometry values are taken from \cite{Til12} and references therein. WISE data in all four bands are also included \citep{Cut12}. The blue line represents the fit based on the best model from \cite{Til12}.  The dashed blue line represents our refined model that better match the CO(3-2) channel maps, with increased scale height and flaring.  For better comparison $\pm$30$\%$ error bars are plotted in red for the photometric measurements between 4.6 and 160$\mu$m.}
\label{SED}
\end{figure}

This set of parameters provides a combination of values that fit reasonably well the observed CO(3-2) channel maps and that are close to the values reported in previous studies.  
The final model keeps matching also the CO radial surface brightness profiles adequately. Nevertheless, when the parameters of the refined model are used to fit the SED (i.e., a thicker disk in both the dust and gas), it is degraded in the mid- and far-infrared. This can be seen in Fig.~\ref{SED}, where a $30\%$ error bars are represented in the photometry measurements from mid- to far-infrared,  which cannot account for the discrepancy observed between model and observations.  
Thus, while the ${\rm CO}(3$--$2)$ channel maps require a thicker gas disk, the mid- and far-infrared part of the SED would require a flatter dust disk.  A possible explanation on this discrepancy is that a significant  amount of dust settling has occurred in the disk, or alternatively that the disk is somehow stratified with a gas layer devoid of dust being present on top of the dust disk, a few scales heights above the mid-plane (the ${\rm CO}(3$--$2)$ $\tau = 1$ surface is located at Z = $33\,{\rm AU}$ at R = $100\,{\rm AU}$).  In order to investigate further these phenomena, theoretical models that accounts for the effects of dust settling and radial migration would be needed. 





\section{Conclusions}
We have presented ALMA observations in ${\rm CO}(3$--$2)$ and continuum at $850\,\mu{\rm m}$. We used these data in combination with MCFOST models to refine the geometrical, physical and chemical  properties of the disk surrounding HD 163296.  The main conclusions are: 
\begin{itemize}

\item For the first time at sub-millimeter frequencies, a detailed disk gas structure is shown, where the front and the back gas disk surface are resolved. 

\item The high spatial resolution and sensitivity of the ALMA data presented in this work required refinements to the model previously presented by \cite{Til12}, from which our model is inspired.  As a result,  the ${\rm CO}(3$--$2)$ channel maps require a thicker gas disk in order to fit the observations, while a flatter dust disk is needed to reproduce the SED in the mid- to far-infrared wavelengths.  We conclude this could be an evidence of the  ${\rm CO}(3$--$2)$  emission comes from gas located further above the disk midplane than the dust, or alternatively an indication of dust settling in the disk. We confirm that the CO freeze out (at $\sim$ 20 K) improves the fit to the CO data as well.

\item We found a clear and strong difference in the radial surface brightness distribution of CO and sub-millimeter dust. A tapered-edge model is unable to fit both sets of data simultaneously with the same density profile for both. The critical radius of the models are similar, but the dust disk requires a much steeper cut-off and sharp outer edge.  We conclude that the adopted tapered edge prescription, valid for a number of disks observed at lower resolution,  must be modified to account for the effects seen with the higher sensitivity and angular resolution now available with ALMA.  We propose a combination of grain growth and inward migration as a plausible explanation of this discrepancy. 
\end{itemize}




\begin{acknowledgements}
 This paper makes use of the following ALMA data: ADS/JAO.ALMA\#2011.0.000010.SV. ALMA is a partnership of ESO (representing its member states), NSF (USA) and NINS (Japan), together with NRC (Canada) and NSC and ASIAA (Taiwan), in cooperation with the Republic of Chile. The Joint ALMA Observatory is operated by ESO, AUI/NRAO and NAOJ. The National Radio Astronomy Observatory is a facility of the National Science Foundation operated under cooperative agreement by Associated Universities, Inc.This publication makes use of data products from the Wide-field Infrared Survey Explorer, which is a joint project of the University of California, Los Angeles, and the Jet Propulsion Laboratory/California Institute of Technology, funded by the National Aeronautics and Space Administration. We are grateful to Dr. Antonella Natta who provided very useful comments. IdG acknowledges the Spanish MINECO grant AYA2011-30228-C03-01 (co-funded with FEDER fund). FMe and AH acknowledges financial support provided by the Milenium Nucleus P10-022-F, funded by the Chilean Government, and by the EU FP7-2011 programme, under Grant Agreement 284405. CP acknowledges funding from the European Commission's 7$^\mathrm{th}$ Framework Program (contract PERG06-GA-2009-256513) and from Agence Nationale pour la Recherche (ANR) of France under contract ANR-2010-JCJC-0504-01

\end{acknowledgements}








\end{document}